\documentclass{webofc}
\usepackage[varg]{txfonts} 
\usepackage[right]{lineno}

\begin{document}

\title{The Rubin Observatory’s Legacy Survey of Space and Time DP0.2 processing campaign at CC-IN2P3}

\author{
        \firstname{Quentin} \lastname{Le Boulc'h}\inst{1}\fnsep\thanks{\email{quentin.leboulch@cc.in2p3.fr}} \and
        \firstname{Fabio} \lastname{Hernandez}\inst{1} \and
        \firstname{Gabriele} \lastname{Mainetti}\inst{1}
}

\institute{CNRS, CC-IN2P3, 21 avenue Pierre de Coubertin, CS70202, 69627 Villeurbanne CEDEX, France}

\abstract{%
        The Vera C. Rubin Observatory, currently in construction in Chile, will start performing the Legacy Survey of Space and Time (LSST) in 2025 for 10 years. Its 8.4-meter telescope will survey the southern sky in less than 4 nights in six optical bands, and repeatedly generate about 2 000 exposures per night, corresponding to a data volume of about 20 TiB every night. Three data facilities are preparing to contribute to the production of the annual data releases: the US Data Facility  will process 35\% of the raw data, the UK data facility will process 25\% of the raw data and the French data facility, operated by CC-IN2P3, will locally process the remaining 40\% of the raw data.

        In the context of the Data Preview 0.2 (DP0.2), the Data Release Production pipelines have been executed on the DC-2 simulated dataset (generated by the Dark Energy Science Collaboration, DESC). This dataset includes 20 000 simulated exposures, representing 300 square degrees of Rubin images with a typical depth of 5 years.

        DP0.2 ran at the Interim Data Facility (based on Google cloud), and the full exercise was independently replicated at CC-IN2P3. During this exercise, 3 PiB of data and more than 200 million files were produced. In this contribution we will present a detailed description of the system that we set up to perform this processing campaign using CC-IN2P3's computing and storage infrastructure. Several topics will be addressed: workflow generation and execution, batch job submission, memory and I/O requirements, etc. We will focus on the issues that arose during this campaign and how we addressed them and will present some perspectives after this exercise.
}
\maketitle
\section{Introduction}
\label{intro}

The Vera C. Rubin Observatory is in the final stages of its construction on the Chilean Andes, and is preparing to execute the Legacy Survey of Space and Time (LSST) that will capture the entire southern sky every four nights in six different bands during ten years. The 8.4-meter telescope together with the 3.2 gigapixels camera (the largest digital camera ever built) will produce tens of petabytes of raw and calibration image data. Data processing pipelines will progressively generate an astronomical catalog of 20 billion galaxies and 17 billion stars and their associated physical properties that will allow scientists to explore the nature of dark matter and dark energy, study the Solar System and the Milky Way, as well as transients objects\,\cite{Ivezic:2019}.

This article presents the Data Preview 0.2 processing campaign performed at CC-IN2P3\footnote{\url{https://cc.in2p3.fr}} for early integration tests of the Rubin data processing pipelines. Section~\ref{DRP} introduces the Rubin data management system and the processing pipelines, section~\ref{processing} details the processing campaign that has been executed, and section~\ref{conclusion} gives some perspectives for future improvements.

\section{The Rubin Data Management System}
\label{DRP}

The Rubin data management system is organized in three main components:
\begin{itemize}
\item the computing, storage and network infrastructure
\item the middleware that provides data access and distributed processing
\item the data processing pipelines
\end{itemize}
In this section we present the middleware components and the data processing pipelines.

\subsection{The Data Butler}
\label{butler}
 
The Data Butler\,\cite{2022SPIE12189E..11J} is the software system that abstracts access to data. Individual sets of data (images and catalog data) are called datasets and are accessible in memory as Python objects through a Python API (a command line interface is also available), without the need to know where the file are located or what file format is used. 

The datasets are organized in a database called the Registry which associates them with their astronomical description (sky coordinates, exposure identification, etc.). The Datastore is the component responsible for the storage of the files associated with the datasets, and provides the read and write interface to Python.

\subsection{The processing pipelines}
\label{pipeline}

Data Release Production will be run every year in order to reprocess the complete set of observations from the beginning of the survey. The following processing operations are performed\,\cite{LDM-151}:
\begin{itemize}
\item Characterization and calibration of individual exposures,
\item Coaddition of single exposures to produce deeper images,
\item Subtraction of images to identify time-variable sources,
\item Processing of the coadded images for object detection and measurement,
\item Measurements on single visit images,
\item Catalog level processing to generate additional object properties.
\end{itemize} 

Data processing pipelines\,\cite{bosch-pipelines} are composed of about 80 distinct processing tasks, implemented in Python with a C++ core.  These tasks are connected through their input and output datasets (mostly catalog and image data), and the resulting Data Release Production processing workflow is represented as a directed acyclic graph (DAG). This large and complex workflow can be split into several sub-workflows called "steps" that correspond to a given level of processing that must be executed in a specific order. Each of these steps is composed of a number of pipeline tasks.

\subsection{The execution framework}
\label{bps}

The pipeline execution framework is called pipetask. It takes as input a description of the pipeline to be executed (list of tasks with their dependencies and configuration), the Data Butler repository where the input data are accessible, and the selection of data to be processed. pipetask provides the ability to execute the corresponding tasks in the correct order, and supports multiprocessing in case that multiple tasks can be executed in parallel when multiple CPU cores are available. It is however restricted to a single node, which is not sufficient for large scale workflows.

\begin{figure}[h]
\centering
\includegraphics[width=6cm]{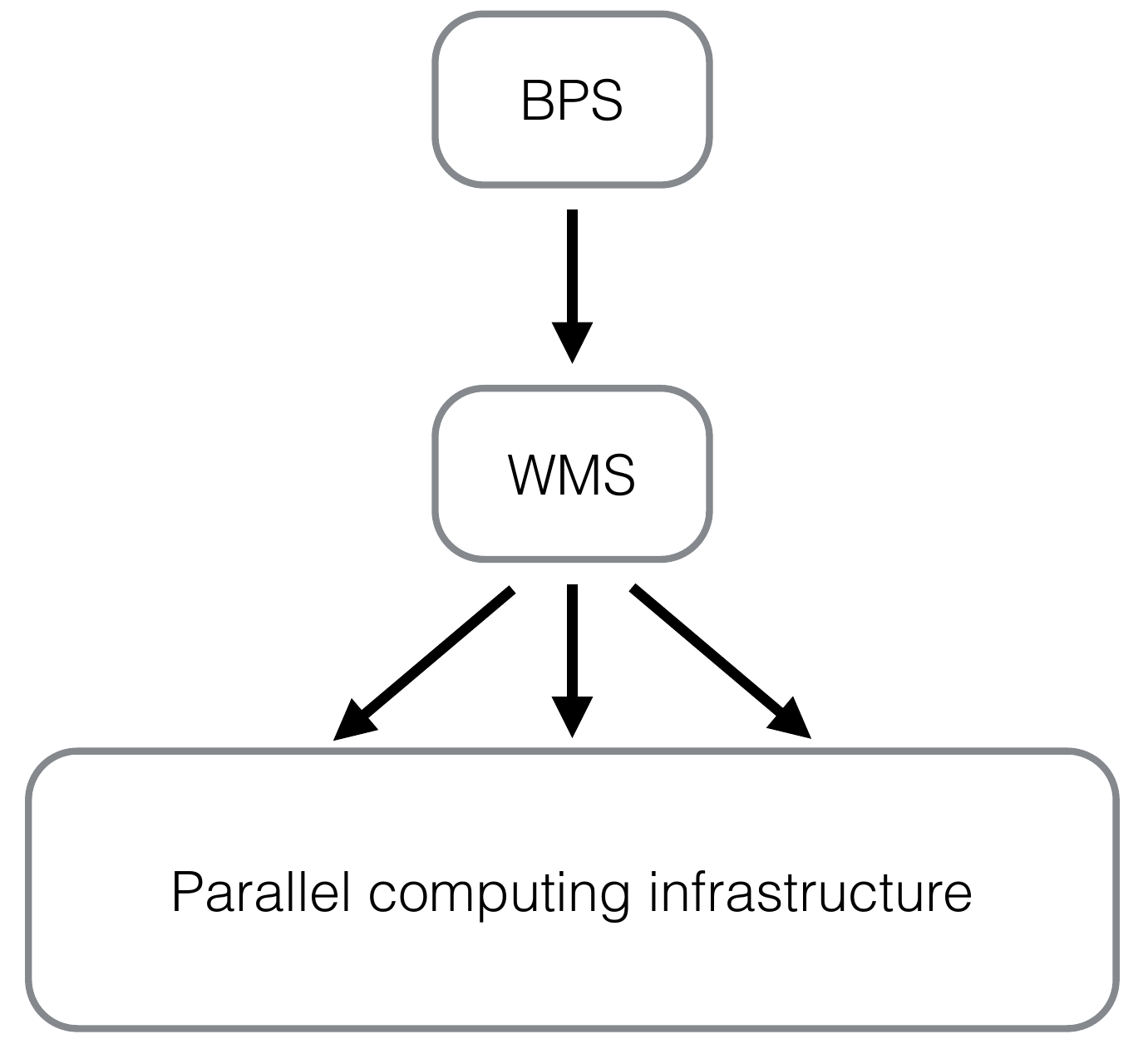}
\caption{Submission of a BPS workflow to a WMS for execution on a parallel computing infrastructure.}
\label{WMS}
\end{figure}

The Batch Processing Service (BPS)\,\cite{bps} is the system for distributed pipeline execution. It provides an additional layer that is able to execute multiple pipetask commands on independent nodes. BPS takes the same input information as pipetask and generates a workflow graph that contains all tasks: commands to be executed, resources requirements, and their dependencies. This workflow graph is then submitted to an external Workflow Management System (WMS) through a dedicated plugin for execution on a parallel computing infrastructure, as shown in figure~\ref{WMS}.

\section{The Data Preview 0.2 processing campaign}
\label{processing}

\subsection{Data Preview 0.2}
\label{DP02}

Data Preview exercises aim at preparing Rubin systems before operations, and help the scientific community to get familiar with Rubin data. In particular, Data Preview 0\,\cite{RTN-001} uses simulated data to perform early integration tests of the processing pipelines and the Rubin Science Platform (RSP)\,\cite{rsp} that provides viewing and analysis tools for images and catalogs. Catalog queries in the RSP are facilitated by Rubin's high performance distributed database Qserv. Data Preview 0 also provides data for preparation of science analysis.

For Data Preview 0.2 (DP0.2)\,\cite{RTN-041}, simulated data provided by the Dark Energy Science Collaboration was reprocessed. These images were generated for their Data Challenge 2\,\cite{lsstdarkenergysciencecollaboration2022desc} and represent 5 years of the LSST wide survey on 300 square degrees. This is approximately  \(0.5 \%\) of the full LSST survey that will cover 18 000 square degrees during 10 years. The DESC DC2 data set includes about 20 000 Rubin camera simulated exposures, each with up to 189 detectors, which makes about 3 million files in total. The total volume of this data set is 50 TiB.

DP0.2 processing was executed at the Interim Data Facility, and it was decided to independently perform the same exercise at the IN2P3 / CNRS Computing Centre (CC-IN2P3) which hosts the French Data Facility. From the processing point of view, the purpose of this exercise was to test the latest versions of both the pipelines and the middleware, check the scalability of the infrastructure, and introduce automation of workflow execution.

After completion of the processing of these simulated data, the resulting data products were made available though the Rubin Science Platform.

\subsection{Workflow execution}
\label{workflow}

The WMS used at CC-IN2P3 for DP0.2 is Parsl\,\cite{babuji1parsl}, a tool that provides parallelism of Python programs on distributed computing resource. It was selected for its scalability, its ability to submit jobs to Slurm\,\cite{slurm} which is the batch system used at CC-IN2P3, and its HighThroughputExecutor features which is suitable for CC-IN2P3 computing infrastructure. In addition, a Parsl plugin\,\cite{parsl_plugin} for BPS was already available and only slight customisations were needed.

The full DP0.2 workflow is too large to be executed at once, so the execution was divided to be executed in several smaller submissions. First, as described in section~\ref{pipeline}, the workflow is divided into several distinct steps that are executed sequentially. In addition, for each step the input data was split into smaller independent groups. The way these groups have been defined takes into account the level of data on which the steps are executed (single exposures, or other partitions of the sky), as well as the resulting size of the generated workflow. On one hand, large workflows require a larger amount of resources to be generated by BPS and longer time to complete. On the other hand, small workflows imply a larger amount of independent submissions to be launched and monitored. Indeed, for each workflow execution it is necessary to identify potential errors during the processing and launch additional submissions to re-execute the missing parts. Finally, the clustering feature of BPS was also used to gather some tasks together in order to reduce the number of tasks to be executed and to improve scalability. The corresponding configuration files are stored in a GitHub repository~\cite{dp02-scripts}.

\subsection{Execution of the processing campaign}
\label{run}

A Data Butler repository for DP0.2 was created and configured, and the input datasets were ingested\,\cite{RTN-029}. The DP0.2 exercise was executed using BPS with the Parsl plugin. The workflow tasks ran within Parsl jobs that submitted to our Slurm computing platform. Before executing the main processing campaign, smaller tests were executed in order to check the Butler repository, the software environment, and progressively increase the number of parallel jobs. During the main campaign, up to 3 000 simultaneous Slurm jobs have been used, and more than $57$ million tasks were executed, using more than $2$ million of computing hours.

\begin{figure}[h]
\centering
\includegraphics[width=9cm]{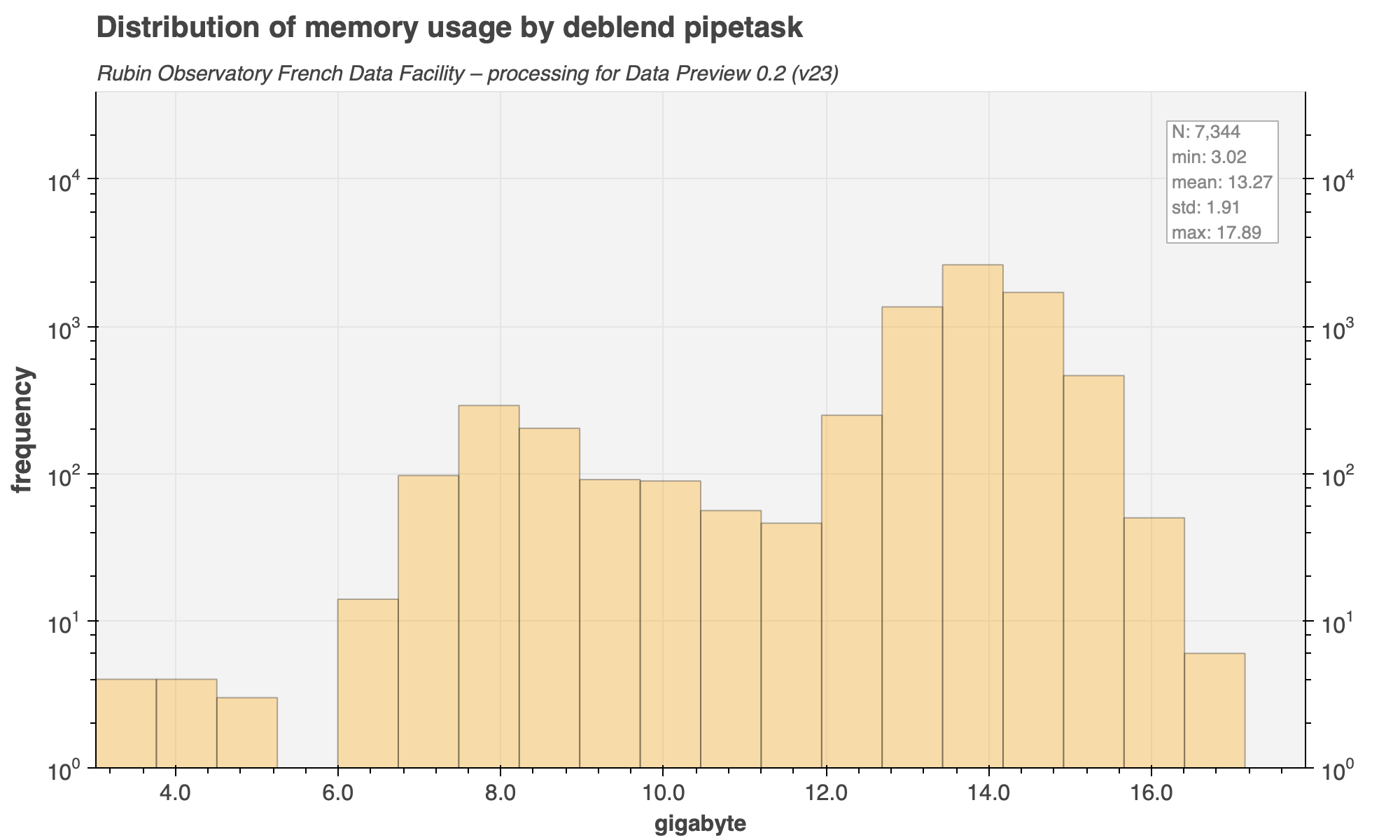}
\caption{Distribution of memory usage for the 7,344 deblend task executions during the DP0.2 processing campaign at CC-IN2P3.}
\label{mem_deblend}
\end{figure}

The tasks were very heterogeneous regarding the type of data they process and produce. The number of times each particular task must be executed varies a lot. In addition the memory and computing resources usage depends a lot on the type of task which is executed, and was not precisely known before launching the processing. Memory usage in particular can vary significantly for a given task with different input data (see figure~\ref{mem_deblend}). It was therefore necessary to continuously monitor the memory usage of the tasks and adjust the jobs configuration accordingly.

Using a custom wrapper script, the tasks were executed within Apptainer\,\cite{apptainer} containers, allowing us to isolate the task environment from the host's. Container images containing the full Rubin software environment including the pipeline software were deployed using CernVM-FS\,\cite{cvmfs}.

The Butler repository Datastore was located on our CephFS\,\cite{cephfs} shared filesystem. $3$ PiB of data  and $201$ M files have been generated during the exercise. The filesystem demonstrated sufficient performance in most of the cases. For one specific type of task we encountered I/O issues and the input files were copied on the local disk of the compute node to improve performance.

We used a PostgreSQL\,\cite{postgresql} database for the Butler repository registry. The database was $314$ GiB in size at the end of the processing. In order to reduce the number of concurrent access, the BPS execution Butler feature\,\cite {DMTN-177} was used: jobs did not access directly the central database but used a local, smaller SQLite\,\cite{sqlite} database instead, where all necessary information was extracted from the Registry beforehand. At the end of the execution, information concerning the new files was ingested into the main Registry.

\subsection{Analysis of the resources usage}
\label{metrics}

An extensive analysis of the processing tasks resource usage is crucial in order to configure properly the Slurm jobs, but also to optimise the pipeline software, and correctly size our infrastructure.

The pipetask framework embeds its own profiling tool that measures the CPU and memory usage of each executed task in the pipeline. These metrics are stored as YAML files in the Butler repository. We retrieved these metrics for all tasks and exported them into CSV files that were analyzed to provide useful graphs.

\begin{figure}[h]
\centering
\includegraphics[width=14cm]{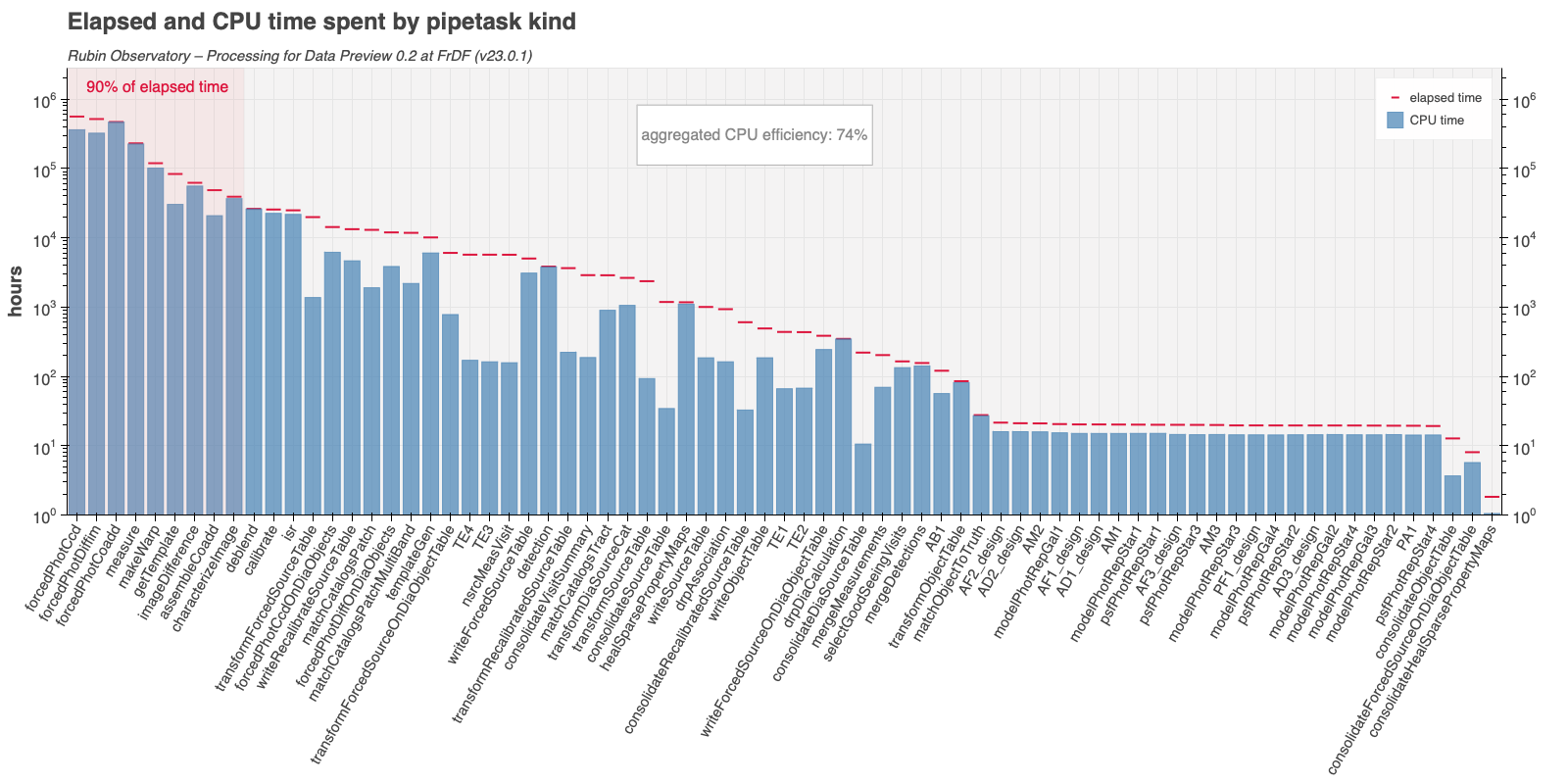}
\caption{Elapsed and CPU time spent for each type of task. Items in the horizontal axis represent the various type of tasks executed in the campaign required to produce a data release, sorted by their aggregated elapsed time.}
\label{cpu}
\end{figure}

Figure~\ref{cpu} shows the walltime and CPU time spent by all type of pipeline tasks. Note that some types of tasks imply multiple execution of the same task on different sets of input data. One can see that $90\%$ of the computing time is spent on the 9 most consuming tasks. In addition, the CPU efficiency\footnote{Understood as the fraction of elapsed time the CPU is actually doing computations, as opposed as waiting for data to process.} for these tasks is rather good, which shows the good performances of our storage system.

\begin{figure}[h]
\centering
\includegraphics[width=14cm]{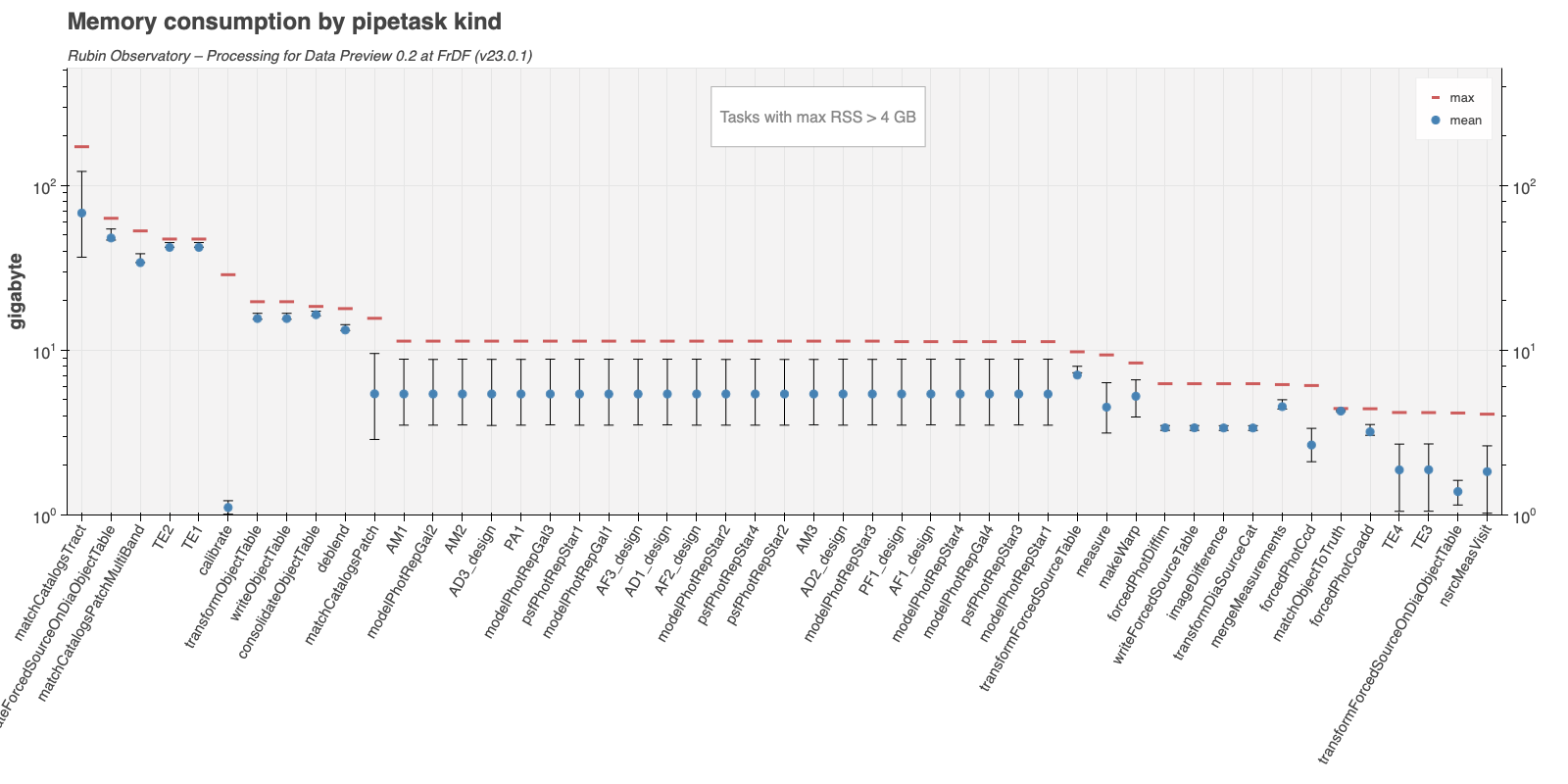}
\caption{Memory consumption for each type of task. Whiskers show the 5th and 95th percentile.}
\label{mem}
\end{figure}

Figure~\ref{mem} shows the distribution of the maximum memory usage per type of task. While most of the tasks have relatively low memory usage in the $1$\,to\,$10$ GiB range, some tasks require several dozen of GiB to run and a few more than $100$ GiB.

\begin{figure}[h]
\centering
\includegraphics[width=9cm]{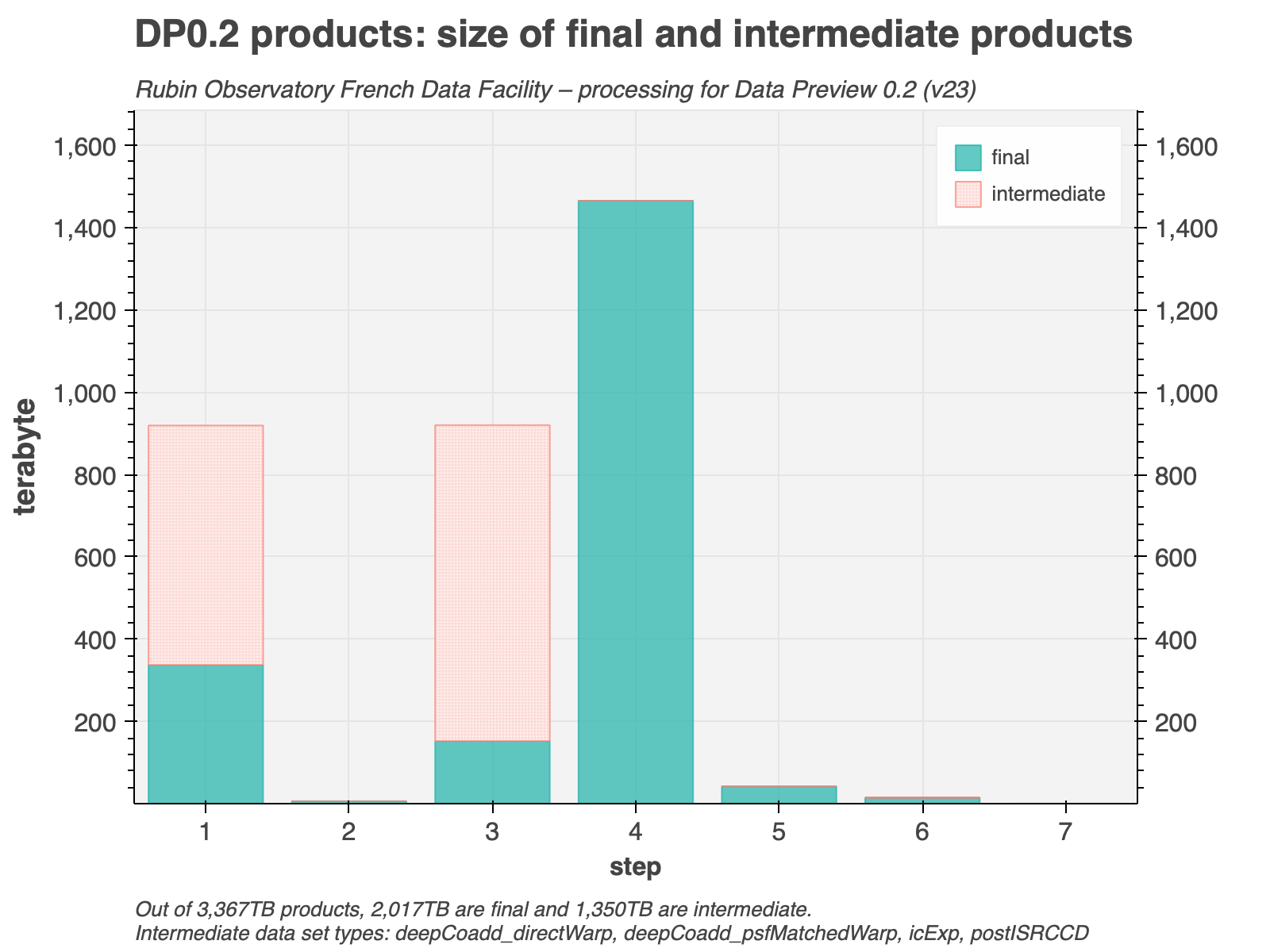}
\caption{Total volume of final and intermediate data products generated by each DP0.2 processing step.}
\label{volume}
\end{figure}

We also collected and analyzed metrics regarding the files and data volume generated by the processing tasks. Figure~\ref{volume} shows the total volume of data products generated by each of the 7 processing steps. Datasets have been tagged as final or intermediate. More than $1$ PiB of data generated in step 1 and step 3 are intermediate files that can be deleted after the processing campaign is complete, however step 4 is the most data volume consuming step and generated more than 1.4 PiB of final products.

\section{Summary and perspectives}
\label{conclusion}

The Vera C. Rubin Observatory is preparing to perform the Legacy Survey of Space and Time. $20$ TiB of raw data will be generated every night for 10 years and the French Data Facility, operated by CC-IN2P3, will contribute resources for processing $40\%$ of the cumulated data. In order to prepare our infrastructure we exercised the Data Release Production pipelines on the DC-2 simulated dataset composed of 20 000 simulated exposures.

This ambitious exercise was successful and showed that the French Data Facility was already able to execute the Rubin data processing pipelines at a significant scale. A similar exercise is ongoing using the dCache\,\cite{dcache} storage system that would allow for a decision on what storage system to use for the operations phase.

One should also note that this processing campaign has been executed fully locally, however multi-site processing is being tested at smaller scale with the PanDA\,\cite{panda-rubin} system. It is also planned to perform more frequent tests with more recent release of the pipelines and of the execution environment. Finally, analysis of the resource usage will be used to understand how the CPU and memory usage of the pipeline can be improved.

\bibliography{references}

\end{document}